# Observation of the Nernst signal generated by fluctuating Cooper pairs

A. Pourret[1], H. Aubin[1], J. Lesueur[1], C. A. Marrache-Kikuchi[2], L. Bergé[2], L. Dumoulin[2], K. Behnia[1]

[1]*Laboratoire de Physique Quantique (CNRS -UPR5), ESPCI, 10 Rue Vauquelin, 75231 Paris, France*

[2]*CSNSM, IN2P3-CNRS Bâtiment 108, 91405 Orsay, France*

**Long-range order is destroyed in a superconductor warmed above its critical temperature ($T_c$). However, amplitude fluctuations of the superconducting order parameter survive[1] and lead to a number of well established phenomena such as paraconductivity[2]: an excess of charge conductivity due to the presence of short-lived Cooper pairs in the normal state. According to an untested theory[3], these pairs generate a transverse thermoelectric (Nernst) signal. In amorphous superconducting films, the lifetime of Cooper pairs exceeds the elastic lifetime of quasi-particles in a wide temperature range above $T_c$; consequently, the Cooper pairs Nernst signal dominate the response of the normal electrons well above $T_c$. In two dimensions, the magnitude of the expected signal depends only on universal constants and the superconducting coherence length, so the theory can be unambiguously tested. Here, we report on the observation of a Nernst signal in such a superconductor traced deep into the normal state. Since the amplitude of this signal is in excellent agreement with the theoretical prediction, the result provides the first unambiguous case for a Nernst effect produced by short-lived Cooper pairs.**

Nernst effect, the generation of a transverse electric field by a longitudinal thermal gradient, has attracted considerable attention since the observation of a puzzling Nernst signal in the normal state of high $T_c$ cuprates[4-10]. In the context of the debate on the origin of this signal, Ussishkin, Sondhi and Huse (USH) calculated the contribution of amplitude fluctuations of the superconducting order parameter on thermoelectric transport[3] and concluded that these fluctuations, described in the Gaussian approximation and responsible for the well-established phenomenon of paraconductivity[2] in the normal state of superconductors, should also generate a Nernst signal. In their model, the main contribution to the Nernst signal comes from the Aslamazov-Larkin term, which represents Cooper pairs with a finite lifetime above Tc[1]. This lifetime decreases with increasing temperature. Therefore, in presence of a thermal gradient, the pairs diffusing towards low temperature live longer than those diffusing towards high temperature and thus the temperature gradient induces a net drift of pair towards low temperature. The deflection of this current by a magnetic field produces a transverse voltage and hence a Nernst effect.

According to USH calculations, the contribution of Gaussian superconducting fluctuations to thermoelectricity leads to a finite off-diagonal component in the Peltier conductivity tensor, $\alpha_{xy}$, which is the ratio of the longitudinal charge current to the transverse thermal gradient ($\alpha_{xy} = J_x/\nabla_y T$). In particular, in two dimensions, and for low magnetic fields $B << \phi_0 / 2\pi\xi^2$, $\alpha_{xy}$ is expected to follow this simple expression:

$$\alpha_{xy}^{SC} = \frac{1}{6\pi} \frac{k_B e}{\hbar} \frac{\xi^2}{\ell_B^2} \qquad (1)$$

Here, $\ell_B = (\hbar/eB)^{1/2}$ is the magnetic length scale. Note that in equation 1, the three universal constants (Planck, Boltzmann and the charge of electron) combine to generate the quantum of thermoelectric conductance ($k_B e / h = 3.3$ $nA/K$), a less celebrated concept compared to the quanta of electric ($e^2 / h$ (ref. 1)) or thermal ($\pi^2 k_B^2 T / 3h$ (ref. 11)) conductance. However, in the notation used by USH (as it is often the case for theoretical papers), $k_B$ is taken as equal to unity and quantum of thermoelectric conductance does not appear explicitly. Since the magnitude of the coherence length $\xi$ is the only parameter in Eq. 1, this theory is particularly apt for an unambiguous confrontation with experiment.

We have tested this theory by measuring the Nernst coefficient in amorphous films of $Nb_xSi_{1-x}$[12,13,14], which present the widely-documented features of superconductor-insulator transition in dirty two-dimensional superconductors[15]. The competition between superconducting and insulating ground states is controlled by the Nb concentration, $x$, the thickness, $d$, or the magnetic field[16].

The data on the two samples presented in Figure 1 and Figure 2 show that a finite measurable Nernst signal persists well above $T_c$. Before comparing this observation with the theoretical prediction by USH, let us argue that neither the excitations of the normal state, nor the superconducting *phase* fluctuations could be a plausible source for the observed Nernst signal. A rough scale for the normal-state Nernst signal is the product of the Seebeck coefficient ($S$) and the Hall angle ($\tan\theta = \dfrac{R_H}{\rho_{xx}}$, with $R_H$ as the Hall coefficient and $\rho_{xx}$ the longitudinal resistivity). As seen in Figure 2, in the entire range of our measurements, the Nernst coefficient, $\nu$, is three orders of magnitude larger than $S\tan\theta$. In a multi-band metal, the contribution of carriers with different signs to $S\tan\theta$ cancel out and its overall value could become smaller than $\nu$ (ref. 17), but such a possibility can be easily ruled out here. The hypothetic existence of two very small Fermi surface pockets hosting carriers of opposite sign with long mean free-path appears implausible. The small value of $\tan\theta \approx 2\times10^{-5}$ simply reflects an extremely short electronic mean-free-path (of the order of inter-atomic distance ~0.25 nm) and a conventional carrier density (the magnitude of $R_H = 4.9\times10^{-11}\ m^3/C$ is comparable to what is reported for bulk Nb (ref. 18).

Can this signal be caused by phase fluctuations of the superconducting order parameter? This is also unlikely. Contrary to the under-doped cuprates, the carrier density in $Nb_{0.15}Si_{0.85}$ is comparable to any conventional metal. Since the "phase stiffness" of a superconductor is determined by its superfluid density[19], there is no reason to speculate on the presence of preformed Cooper pairs without phase coherence in a wide temperature window above $T_c$ as it has been the case in the pseudo-gap state of the cuprates. In contrast to granular superconductors[20], decreasing the thickness leads to a shift of the sharp superconducting transition and does not reveal a temperature scale other than the mean-field BCS critical temperature. The variation of $T_c$ with thickness has been attributed to the enhancement of the Coulomb interactions with the increase in the sheet resistance, $R_{square}$[21].

On the other hand, there is no reason to doubt the presence of amplitude fluctuations of the superconducting order parameter invoked by the USH theory. Now, the theory makes a precise prediction on the magnitude of $\alpha_{xy}$, but what is directly measured by the experiment is the Nernst coefficient, $\nu$, which is intimately related to it. When the Hall angle is small and the contribution of superconducting fluctuations to charge conductivity is also small, there is a simple relationship between $\alpha_{xy}$, $\nu$, and the sheet resistance $R_{square}$:

$$\frac{\alpha_{xy}^{SC}}{B} = \nu \sigma_{xx} = \frac{\nu}{R_{square}} \qquad (2)$$

The validity of both conditions was checked in our experiment: $\tan \theta \approx 2 \times 10^{-5}$ and $\sigma^{SC} = \frac{e^2}{16\hbar} \frac{T_c}{T - T_c}$ (ref. 1) is a few percent of $\sigma_{xx}$ when T>1.1$T_c$.

Because $\alpha_{xy}^{SC} \propto B$, it follows from the relation 2 that $\nu$ should be independent of the magnetic field, in the low magnetic fields region considered by this model. The data show that this is indeed the case. Therefore, we can directly determine $\alpha_{xy}/B$ for each temperature in the zero-field limit using equation 2 and the data for $\nu$ and $R_{square}$. Now, equation 1 can be rewritten as:

$$\xi^2 = \frac{6\pi\hbar^2}{k_B e^2} \frac{\alpha_{xy}}{B} = 5.9 \times 10^{-7} \frac{\alpha_{xy}}{B} \qquad (3)$$

The value of $\xi$ obtained in this way for the two samples is displayed in Figure 3 as a function of the reduced temperature $\varepsilon = (T-T_c)/T_c$ and allows a direct verification of the theory. Theoretically, the coherence length, $\xi$, should vary as $\varepsilon^{-1/2}$ (ref. 1). Moreover, its absolute magnitude is expected to scale inversely with $\sqrt{T_c}$ which is in conformity with what is found here: The ratio of the coherence lengths $\xi_1$ and $\xi_2$ for samples 1 and 2 is $\xi_1(\varepsilon=1)/\xi_2(\varepsilon=1) = 1.48$, while the ratio $\sqrt{(T_{c2}/T_{c1})} = 1.52$. More quantitatively, the coherence length of a two-dimensional dirty superconductor is[1]:

$$\xi_d = \frac{1}{\sqrt{\varepsilon}} 0.36 \sqrt{\frac{3}{2} \frac{\hbar v_F \ell}{k_B T_c}} \qquad (4)$$

Where $v_F$ and $\ell$ are the Fermi velocity and the electronic mean-free-path. The most direct way to estimate $v_F \ell$ is to use the known values of the electric conductivity, $\sigma \approx 6.4 \times 10^4 \ \Omega^{-1} cm^{-1}$ and the electronic specific heat, $\gamma_e \approx 108 \ JK^{-1}m^{-3}$ (see methods). The generic relationship between the specific heat and the thermal conductivity ($\kappa$), combined with the Wiedemann-Franz law yields:

$$v_F \ell = 3 \frac{\kappa}{\gamma_e T} = \left(\frac{\pi k_B}{e}\right)^2 \frac{\sigma}{\gamma_e} \qquad (5)$$

This allows us to directly estimate $v_F \ell = 4.35 \times 10^{-5} \ m^2 s^{-1}$ and, using equation 4, plot $\xi_d(\varepsilon)$. As seen in the figure, for both samples, for small values of $\varepsilon$, there is an excellent agreement between these two estimations of the coherence length. As the temperature rises, $\sqrt{\alpha_{xy}}$ decreases faster than $\xi_d$. This discrepancy is not surprising since, the $\varepsilon^{-1/2}$ dependence of the coherence length $\xi_d(\varepsilon)$ and the USH theory are valid only for small(<1) values of $\varepsilon$. Moreover, $\xi$ becomes much smaller than the film thickness and the 2D limit is no more valid. It is remarkable, however, that even for $\varepsilon$=10, the two values obtained for $\xi$ differ by a mere factor of two.

In retrospect, it is not surprising that this effect is unambiguously observed for the first time in a dirty superconductor. According to the theory, what is universal is the magnitude of $\alpha_{xy}^{SC}$, the expected Nernst signal is therefore larger when the normal-state conductivity is lower. Moreover, due to the short mean-free-path of electrons, the normal-state Nernst effect becomes negligible, making the detection of the signal produced by superconducting fluctuations easier. On a more fundamental level, in a dirty superconductor, the lifetime of a fluctuating Cooper pair ($\tau_{GL} \approx \xi^2 / v_F \ell$ (ref. 1)) exceeds the elastic lifetime of a normal electron ($\tau_{el} \approx \ell / v_F$) in a wide temperature window above $T_c$, paving the way for a dominant contribution of the Gaussian fluctuations to the Nernst signal.

Charge conductivity, even in presence of Gaussian fluctuations, is dominated by the contribution of normal electrons. As we saw above, this is not the case for the Nernst effect, which (due to the smallness of the normal-state Nernst effect) can be totally dominated by these fluctuations. This makes the Nernst effect a powerful probe of superconducting fluctuations.

We conclude by considering the field-dependence of the Nernst coefficient. The USH calculation has been performed for weak fields ($\xi << \ell_B$) and was only tested here in the zero field limit. For T>$T_c$, as seen in Fig. 1, the Nernst signal reveals a field scale in its field dependence which increases with increasing temperature. This appears to simply reflect the decrease in the field scale associated with $\xi(B^* = \hbar/(e\xi^2))$. For both samples, the Nernst signal does not vanish even with the application of a field as large as 4 T. This field is larger than all three field scales one can associate with the destruction of superconductivity. These are i) The critical field for the Superconductor-Insulator transition defined as the crossing field of R(B) curves at low temperatures, $B^{SI}$ (ref. 16); ii) The Pauli limit $B^{P}$ =1.84Tc (ref. 22); and iii) the orbital limit $B_{orb} = \phi_0 / 2\pi\xi^2$. The values for both samples are shown in the table 1, where it appears that the upper critical field $B_{c2}$ is set by Pauli limit, i.e. $B_{c2} = B^{SI} \approx B^{P} < B^{orb}$. It is natural to assume that the superconducting long-range order is indeed destroyed at $B_{c2}$, but the *superconducting fluctuations* persist and gradually fade away above $B_{c2}$, as they do above $T_c$. The contribution of the Gaussian fluctuations to the Nernst effect in high magnetic fields remains a challenging question for both theory and experiment.

**Methods**

The two amorphous thin films of $Nb_{0.15}Si_{0.85}$ used in this study were prepared as described elsewhere[12,13]. The nominal concentration of Nb in the two samples used in this study was the same (x=0.15) and the difference in $T_c$s, (0.165 K in sample 1 with d=12.5 nm and 0.38 K in sample 2 with d=35 nm), is mainly due to the difference in the thickness of the two samples. The critical temperature was defined as the mid-height of the resistive transition at zero field. A set-up with one resistive heater, two $RuO_2$ thermometers and two lateral contacts[23], was used in order to measure the thermoelectric and the electric coefficients of each sample in a dilution cryostat. At T~0.19 K, we could resolve a DC voltage of 1 nV and a temperature difference of 0.1 mK. The magnitude of the electronic specific heat used for the estimation of the coherence length is based on the magnitude of $\gamma_e$ in bulk Nb (ref. 24) and the concentration of itinerant electrons provided by Nb fraction, as confirmed by direct measurements of specific heat in $Nb_xSi_{1-x}$ thin films of a lower concentration[13].

**Acknowledgments** This work is supported by Agence Nationale de Recherche.

**Competing interests statement** The authors declare that they have no competing financial interests.

**Correspondence** and requests for materials should be addressed to H.A. (Herve.Aubin@espci.fr) and/or

K.B. (Kamran.Behnia@espci.fr)

**Table 1 The samples and their parameters**

| Sample | d[nm] | $T_c$[K] | $\xi_d(\varepsilon=1)$[nm] | $B^{SI}$[T] | $B^P$[T] | $B^{orb}$[T] |
|--------|-------|----------|----------------------------|-------------|----------|--------------|
| 1 | 12.5 | 0.165 | 19.7 | 0.36 | 0.3 | 0.85 |
| 2 | 35 | 0.38 | 13 | 0.91 | 0.7 | 1.95 |

$T_c$ is defined as the temperature at which the resistance is half the normal state value. For the determination of $\xi_d$, see text. $B^{SI}$ is the critical field associated with the superconductor-insulator transition, $B^P$ and $B^{orb}$ are respectively the Pauli and the orbital limiting fields.

**Figure 1: Nernst signal from sample 1**

Panels a) and b) : The Nernst signal (N) as function of magnetic field for temperatures ranging from 0.19 K up to 5.8 K, for sample 1 with $T_c$=0.165 K as detected by its resistive transition. A finite Nernst signal is present for T > Tc. With increasing temperature, this signal decreases in magnitude and becomes more field-linear. The panel c) presents the Nernst coefficient, $v = N/B$, for the same sample as a function of magnetic field in a log-log scale. Note that, save for the lowest temperatures, the Nernst coefficient is constant at low magnetic field.

**Figure 2: Nernst signal from sample 2**

Panel c) : The evolution of the Nernst signal with temperature in sample 2 on a semi-log plot. The red curve marks the onset of superconductivity. Note the evolution of the Nernst signal across the critical temperature. The large Nernst signal below $T_c$ is caused by vortex movement due to the thermal gradient and the reduction of the signal at lower fields for T=0.25 K is due to vortex pinning in the low-temperature-low-field region of the (B,T) plane. Upper panels show the temperature dependence of the resistivity, panel b),  and the Nernst coefficient, panel a). The Nernst coefficient, which exceeds the measured value of *Stanθ* at 2 T multiplied by 2000, cannot be attributed to the normal-state quasi-particles.

**Figure 3: Temperature dependence of the coherence length**

 The coherence length, ξ, directly deduced from $\alpha_{xy}(\xi^2 = \dfrac{6\pi\hbar^2}{k_B e^2}\dfrac{\alpha_{xy}}{B})$ for the two samples as function of reduced temperature, ε. The solid lines represent $\xi_d = \dfrac{1}{\sqrt{\varepsilon}}0.36\sqrt{\dfrac{3}{2}\dfrac{\hbar v_F \ell}{k_b T_c}}$ for each sample. The agreement between these two estimations of the coherence length for small ε provides compelling evidence for the validity of the Gaussian fluctuations as the source of the Nernst signal.

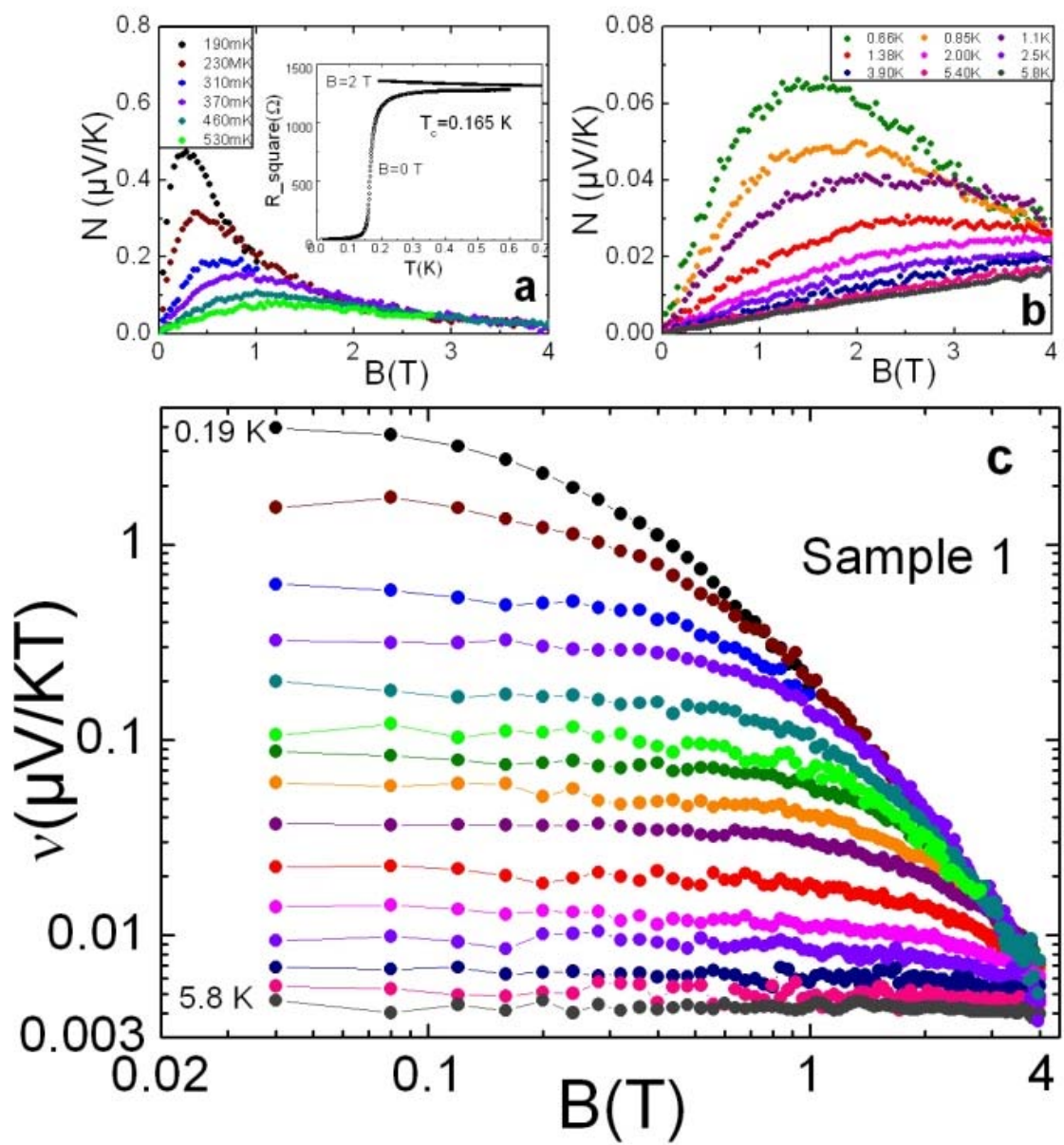

**Figure 1.**

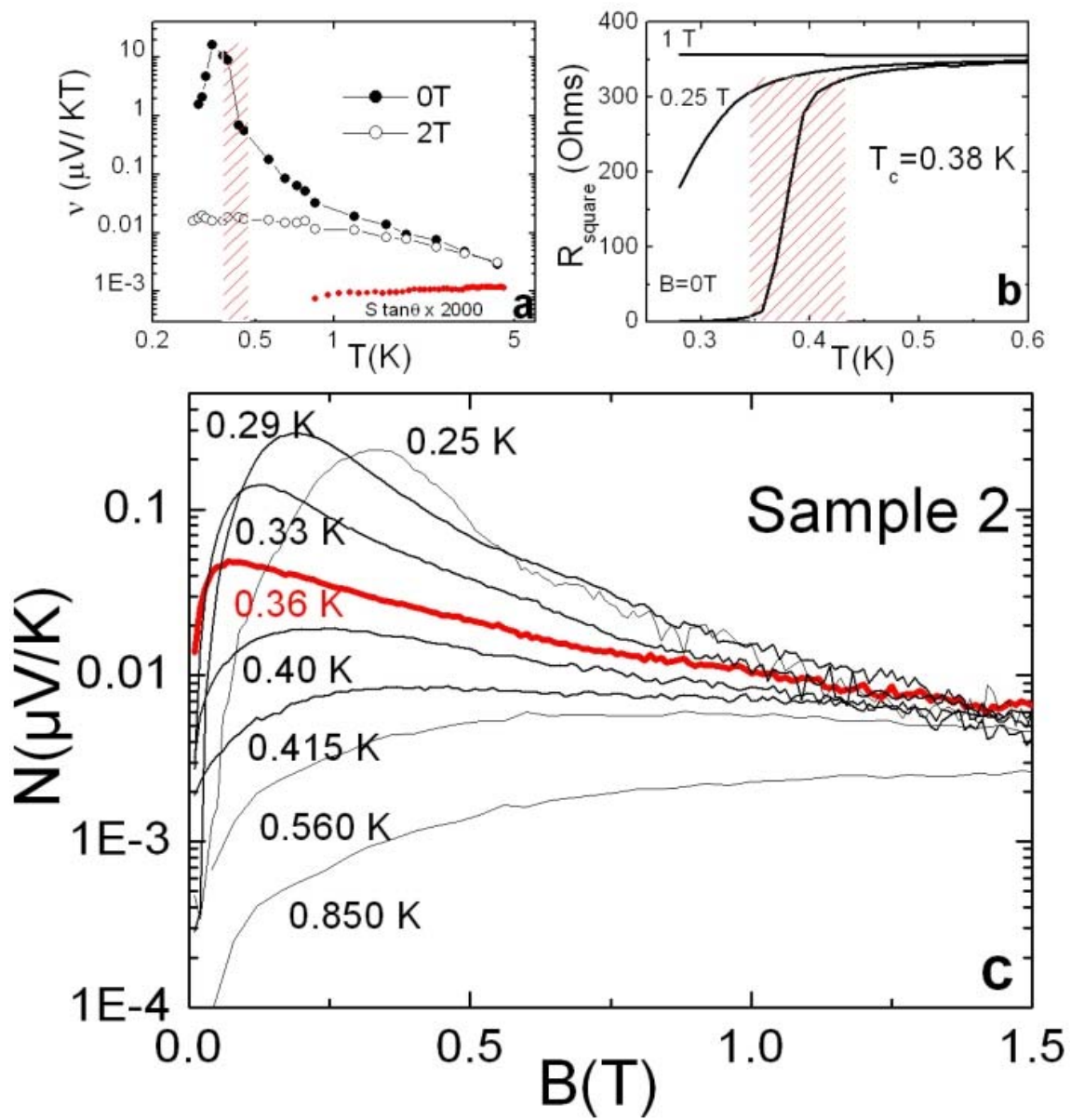

**Figure 2.**

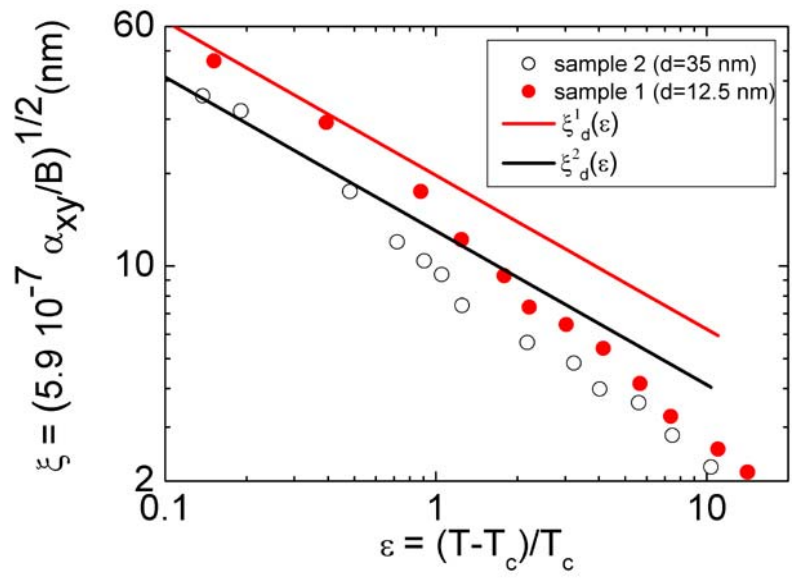

**Figure 3.**